\documentclass[aps,prl,reprint,superscriptaddress,showpacs]{revtex4-1}

\usepackage{graphicx}
\usepackage{dcolumn}
\usepackage{bm}
\usepackage{amssymb}
\usepackage{mathrsfs}

\begin{document}

\title{Direct Observation of Instantaneous Influence between Entangled Photons}


\author{Boya Xie}
\affiliation{
Hubei Key Laboratory of Modern Manufacturing Quantity Engineering, School of Mechanical Engineering, Hubei University of Technology, Wuhan, Hubei, P.R.China, 430068}
\author{Peng Yang}
\affiliation{School of Electrical and Electronic Information Engineering, Hubei Polytechnic University, Huangshi, Hubei 435003, P.R. China}
\author{Guanyang Zhang}
\author{Lei Nie}
\author{Zhongsheng Zhai}
\author{Xuanze Wang}
\author{Sheng Feng}
\email[]{fengsf2a@hust.edu.cn}
\affiliation{
Hubei Key Laboratory of Modern Manufacturing Quantity Engineering, School of Mechanical Engineering, Hubei University of Technology, Wuhan, Hubei, P.R.China, 430068}


\date{\today}

\begin{abstract}
We investigate direct observation of quantum nonlocality without reference to theoretical models (including Bell theorem) except quantum mechanics, with a bipartite polarization-entangled state in which one photon immediately reduces into a circular-polarization (CP) state after its partner is detected in another CP state. Of essence is the mechanical detection of the CP state of a photon that carries angular momentum and exerts a torque on a half-wave plate whose mechanical motion is then varied. If implemented, the model-independent observation of quantum nonlocality violates Lorentz invariance in experiment and may indicate new fundamental physics beyond the Standard Model.
\end{abstract}


\maketitle


Eight decades ago, one of the most far-reaching debates in scientific history occurred between Bohr and Einstein on the completeness of quantum mechanical description of physical reality \cite{Einstein1935,Bohr1935}. It was the statistical nature of quantum mechanics (QM) that first caught the attention of Einstein who believed that quantum systems were controlled by hidden variables that determined measurement outcomes. Later in 1935, Einstein, Podolsky, and Rosen (EPR) brought the nonlocal feature of QM into the public sight \cite{Einstein1935}, stimulating many attempts \cite{Bohm1952,Bohm1957,Bell1964,Clauser1969,Pitowsky1982,Greenberger1990,Hardy1993,Wiseman2007} to explore the EPR paradox for a satisfactory solution.

A turning point of the long-lasting debate came in 1964 when Bell published an inequality \cite{Bell1964} giving an upper bound on the strength of some correlations exhibited by local realist theories. According to Bell theorem, violation of the inequality would conclusively preclude all local realistic theories. But early Bell experiments \cite{Freedman1972,Aspect1982} were performed under imperfect conditions and forced to make additional assumptions to deny local realism \cite{Genovese2005}. Long after those pioneering works, three ``loophole-free" Bell experiments were reported at last in 2015 \cite{Giustina2015,Shalm2015,Hensen2015}.

Bell theorem, together with Bell experiments, has deeply influenced our perception and understanding of physics and is essential for the applications of quantum information technologies. Quantum nonlocal correlations, as witnessed by the violation of Bell inequality, are now thought as fundamental aspects of quantum theory by many physicists \cite{Walborn2011,Cavalcanti2011,Christensen2013,Hirsch2013,Erven2014,Brunner2014,Popescu2014,Aspect2015}. Yet, complete consensus has not been reached in the literature over whether the door has been closed on the Bohr-Einstein debate \cite{Aspect2015,Khrennikov2015,Kupczynski2017,Pons2017}. In the spectrum of opinions on the results of Bell experiments, still on the focus of the question is quantum nonlocality whereby particles appear to influence one another instantaneously \cite{Orlov2002,Matzkin2008,Khrennikov2015,Kupczynski2017,Pons2017}.

\begin{figure}[htbp]
\centering
\includegraphics[width=7cm]{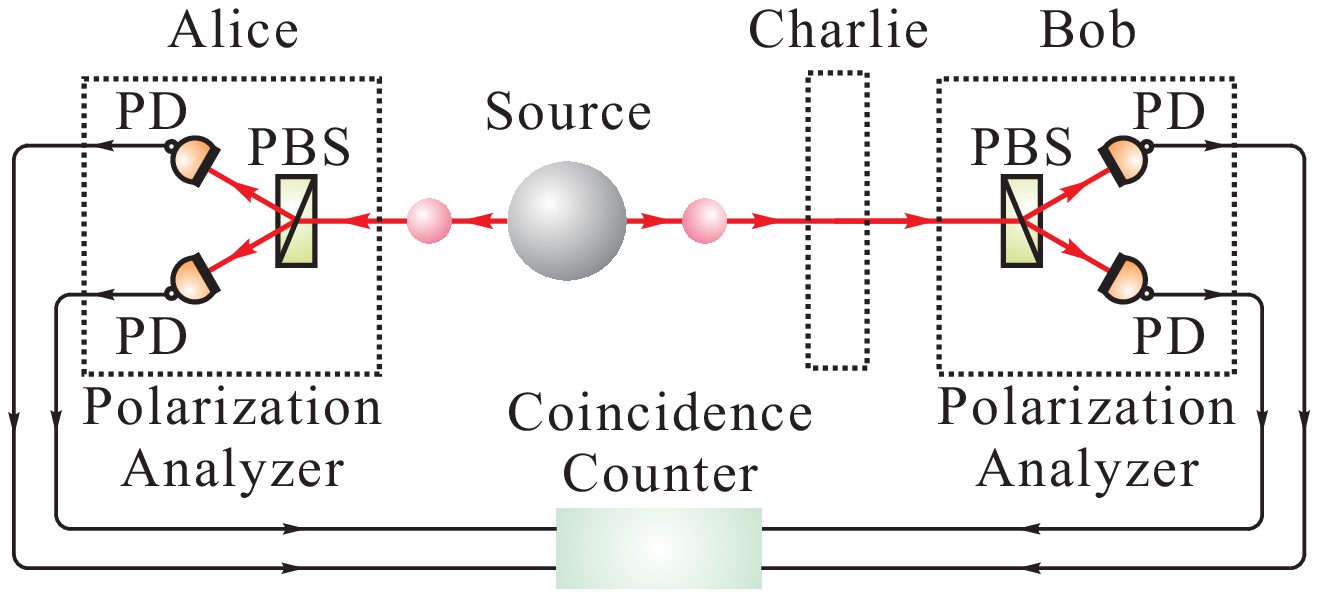}
\caption{(color online)  A typical Bell experiment on nonlocal correlations between entangled photons, which is witnessed by the correlations of the local measurement results of Alice and Bob who are widely separated in space. PD: Photon detector. PBS: Polarizing Beamsplitter.}
\label{fig:bell}
\end{figure}

In a typical configuration of Bell experiment, quantum nonlocality is usually demonstrated with polarization entangled photons as shown in Fig. \ref{fig:bell}. Suppose that the two photons shared by Alice and Bob are in an entangled state described by,
\begin{eqnarray}\label{eq:ent1}
|\psi\rangle&=&2^{-1/2}(|H\rangle_1|H\rangle_2+|V\rangle_1|V\rangle_2),
\end{eqnarray}
a superposition of $|H\rangle_1|H\rangle_2$ (both photons horizontally polarized) and $|V\rangle_1|V\rangle_2$ (both vertically polarized). According to QM, the polarization states of the two photons are undetermined unless one of them is measured by a polarization detector and irreversibly projected into an eigenstate of an operator describing the measurement. If Alice's photon is in $|H\rangle$ (or $|V\rangle$) state after the measurement, it follows from Eq. (\ref{eq:ent1}) that then Bob's photon will immediately collapse into $|H\rangle$ (or $|V\rangle$) state too, or vice versa, due to the instantaneous influence between the two photons no matter how widely they are separated in space. This nonlocality is referred to as ``spooky action at a distance" by Einstein.

It is Bell theorem, as a widely-accepted theoretical model, that bridges the results of Bell experiments and the nonlocal nature of a complex system described by for example Eq. (\ref{eq:ent1}), which contains no information of spatiotemporal coordinates. Bell experiments share two features with all later investigations on quantum nonlocality: (1) Model dependence and (2) Lorentz invariance \cite{Hardy1992}. The model dependence comes from the fact that all relevant studies must first assume a local realist model that is then rejected according to some criteria \cite{Bell1964,Greenberger1990,Hardy1993,Wiseman2007}, whereas the feature of Lorentz invariance is partially related to the lack of spatiotemporal coordinates in the wave function of the complex system.

In this letter, however, we will show that it is possible to directly observe the instantaneous influence between entangled photons in an experiment, which assumes no local realist theory and hence is model independent. The model independence guarantees that the experiment, if successfully conducted, will be a more convincing test of quantum nonlocality. In addition, the result of the experiment is sensitive to the inertial reference frame where the observation is performed, i.e., it violates Lorentz invariance.

Violation of Lorentz invariance (VLI) has been suggested by a number of theories towards unifying gravity with other fundamental interactions \cite{Kostelecky2011,Pospelov2012}. VLI as a possible and central deviation from known physics is considered as a result of suppressed relics from Planck-scale physics. Because the highest-energy experiments or observations are well below the Planck scale and it is suggested that VLI may be detectable via precision measurements at lower energy level \cite{Kostelecky1995}, widespread interests have been stirred in experimental searching for VLI with cold atoms or ions \cite{Wolf2006,Dzuba2016,Bars2017}. Although these unprecedented experiments have lead to significantly suppressed upper limits on the parameters characterizing VLI in the Standard Model extension \cite{Kostelecky2011}, no positive VLI signals have been identified. Here we will show that VLI is present in direct observation of quantum nonlocality and should be observable with currently available technologies.

\begin{figure}[htbp]
\centering
\includegraphics[width=7cm]{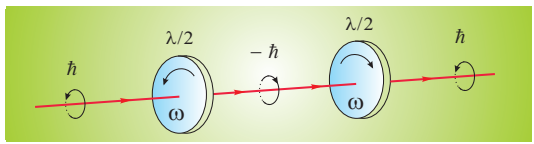}
\caption{(color online)  A mechanical detector for detection of the CP state of a photon, which changes the rotational speed of each half-wave plate by $\omega=\pm 2\hbar/I$ ($I$ is the moment of inertia) \cite{Beth1936}, with the sign of $\omega$ depending on the CP state of the incident photon. In contrast, a photon in a linear-polarization (LP), i.e., $|H\rangle$ or $|V\rangle$, state carries no angular momentum and cannot change the mechanical motion of the plate. One should stress that a CP photon remains in the same CP state as the input after exiting from the detector.}
\label{fig:polana}
\end{figure}

To begin with, let rewrite Eq. (\ref{eq:ent1}) as,
\begin{eqnarray}\label{eq:angular}
|\psi\rangle&=&2^{-1/2}(|L\rangle_1|R\rangle_2+|R\rangle_1|L\rangle_2),
\end{eqnarray}
in which $|L\rangle=(|H\rangle-i|V\rangle)/\sqrt{2}$ and $|R\rangle=(|H\rangle+i|V\rangle)/\sqrt{2}$ stand for the left-hand and right-hand CP states, respectively. It follows from Eq. (\ref{eq:angular}) that, when Alice's photon is detected in $|L\rangle$ (or $|R\rangle$) state, Bob's photon will collapse immediately into $|R\rangle$ (or $|L\rangle$) state \cite{Pan2000} and then carry angular momentum \cite{Beth1936}, or vice versa. Of essence in direct observation of quantum nonlocality is that Bob's photon carrying angular momentum can exert a mechanical torque on a half-wave plate \cite{Beth1936}. Based on the physical process of angular momentum exchange between a half-wave plate and a CP photon, one may construct a mechanical detector consisting of two freely-rotating half-wave plates in a row (Fig. \ref{fig:polana}) to detect, but not to alter, the polarization state of a photon: After the CP photon exerts a torque on a plate when passing through it, the angular momentum of the plate will change by twice that of the incident photon, i.e., $|l_i|=2\hbar$ ($i=1,2$ are the sequential numbers for the plates and $\hbar\equiv h/2\pi$, where $h$ is Planck constant), but the signs of $l_i$ are opposite. At the output of the second plate, the CP state of the photon remains the same as the input of the first plate, but the CP state information of the photon is recorded by the mechanical detector that outputs a corresponding value of $\Delta l=\pm4\hbar$ ($\Delta l \equiv l_2-l_1$). The sign of $\Delta l$ reveals the specific CP state of Bob's photon, into which the photon collapses as a result of detecting Alice's photon in another CP state at a space-like distance.

The core elements in direct observation of instantaneous influence between entangled photons can be fully described in the theoretical framework of QM: (a) Bob's photon is immediately projected into a CP state by the detection of Alice's photon in another CP state according to Eq. (\ref{eq:angular}) \cite{Pan2000}; (b) A photon in a CP state changes the angular motion of a half-wave plate by exerting a mechanical torque on it \cite{Beth1936}. Therefore, the expected results if achieved successfully will be a model-independent evidence for quantum nonlocality since no local realist model is needed.


In connection to the Bell experiment as depicted in Fig. \ref{fig:bell}, someone (Charlie) may place a mechanical detector as shown in Fig. \ref{fig:polana} in front of Bob's state analyzer. Since Charlie's detector does not change the polarization state of Bob's photon, neither Alice nor Bob would be able to know whether Charlie's detector is present in the experiment. However, the output of Charlie's detector is sensitive to the timing sequence of the experiment: If Alice receives a photon before Charlie does, his photon will be reduced into a CP state before arriving at the mechanical detector whose output will be $\Delta l=\pm4\hbar$ with the sign depending on the CP state of the photon detected by Alice. Otherwise, if Alice receives a photon after Charlie, the polarization state of his photon will be undetermined at the input of the mechanical detector, whose output will then be $\Delta l=0$.

In the following, we will show that the sensitivity of Charlie's detector to the stated timing sequential leads to violation of Lorentz invariance. To see this, let consider the following Lorentz transform,
\begin{eqnarray}\label{eq:lorentz}
&&x'_i=\frac{x_i-vt_i}{\sqrt{1-v^2/c^2}},\ \ 
t'_i=\frac{t_i-vx_i/c^2}{\sqrt{1-v^2/c^2}},
\end{eqnarray}
wherein $(x_i,t_i)$ are the spatiotemporal coordinates for Alice's detector $(i=1)$, Bob's detector $(i=2)$, and Charlie's detector $(i=3)$ in the lab reference frame, and $(x'_i,t'_i)$ are the corresponding coordinates in a moving reference frame at the velocity of $v$ along the photon propagation direction ($x$-axis). From Eq. (\ref{eq:lorentz}), it follows that
\begin{eqnarray}\label{eq:timing}
t'_3-t'_1&=&\frac{(t_3-t_1)-v(x_3-x_1)/c^2}{\sqrt{1-v^2/c^2}},
\end{eqnarray}
which shows that the sign of $t'_3-t'_1$ is determined by the direction of $v$ provided $|(t_3-t_1)|<|v(x_3-x_1)/c^2|$. In other words, the timing sequence of ($t'_1, t'_3$) could be different conditioned on the moving direction of the reference frame in which the observation is performed.

Suppose an observer, David, is moving at a constant velocity of $v>0$ such that
\begin{eqnarray}\label{eq:VLIcon}
|v|&>&c^2|t_3-t_1|/|x_3-x_1|,
\end{eqnarray}
then $t'_3-t'_1<0$ according to Eq. (\ref{eq:timing}), i.e., Alice will receive a photon later than Charlie does in David's inertial frame. So, the state of the incident photon into Charlie's detector is undermined and each plate gains no angular momentum. In this case, David will observe $\Delta l=0$ at the output of Charlie's detector. However, if there is another observer, say Frank, moving at $v<0$ satisfying the inequality condition (\ref{eq:VLIcon}), then $t'_3-t'_1>0$ according to Eq. (\ref{eq:timing}), i.e., Alice will receive her photon earlier than Charlie in Frank's inertial frame. So, the photon flying towards Charlie's detector will collapse into a CP state as a result of the instantaneous influence of detecting Alice's photon in another CP state. In this case, Frank will observe $\Delta l =\pm4\hbar$ with the sign relying on the CP state of Alice's photon, accompanied by a variance of the mechanical motion of each plate in Charlie's detector.

In Frank's frame, there is a cause-effect relationship between the local measurements by Alice and Charlie at space-like distance, which is a direct consequence of quantum nonlocality, i.e., the faster-than-light influence between the entangled photons: The detection of Alice's photon in a CP state reduces instantaneously Charlie's photon into another CP state before it enters into his detector. But this cause-effect relationship is not present in David's frame, where Charlie receives a photon earlier than Alice but her photon is not affected by Charlie's measurement in which his photon does not undergo state reduction. That the stated cause-effect relationship in Frank's frame is absent in David's frame violates Lorentz invariance. 

Direct observation of quantum nonlocality and VLI should bring a substantial advance to the study of fundamental physics beyond the Standard Model \cite{Kostelecky2011}. However, the relevant physical effect may be too small to observe: To monitor the angular momentum change of a wave plate caused by a single photon is technically intractable. In what follows, we will discuss how to implement the stated observation in experiment with currently available technologies. Firstly, one should note that many photons per second can be generated with regular photon-pair sources, facilitating the observation of the expected physical effect. Secondly, in light of the technical challenge in experiment, one may utilize an optical amplifier to boost the power of the photon-bearing beam before it enters into Charlie's detector.

\begin{figure}[htbp]
\centering
\includegraphics[width=8cm]{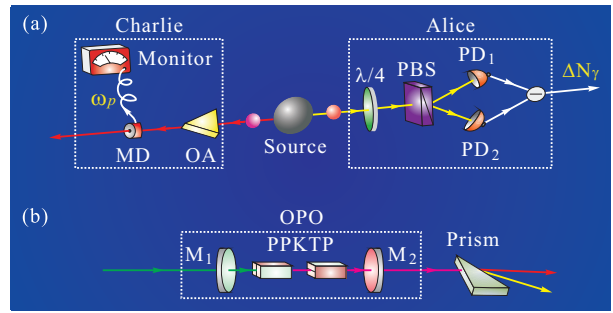}
\caption{(color online) (a) Schematics for direct observation of instantaneous influence between entangled photons, through mechanical detection of the angular momenta of photons. MD: Mechanical detector. OA: Optical amplifier. $\lambda/4$: 1/4-wave plate. PBS: Polarizing beamsplitter. PD$_{1,2}$: Photon detectors. (b) A type-I frequency-nondegenerate optical parametric oscillator (OPO) as a bright source for polarization-entangled photon-pair generation. Inside the OPO cavity, one PPKTP crystal is rotated by 90$^\circ$ around the beam direction relative to the other crystal. M$_{1,2}$: Cavity mirrors. PPKTP: Periodically-poled Potassium Titanyl Phosphate (KTP).}
\label{fig:exp1}
\end{figure}

Let suppose that one has a photon-pair source emitting $N_\gamma$ pairs per second in the state of Eq.(\ref{eq:angular}) and an optical amplifier with a power gain of $G$ (Fig. \ref{fig:exp1}a). One photon (Alice's photon) of each pair is measured by a CP analyzer and projected into $|L\rangle$ or $|R\rangle$ state. Immediately, the other (Charlie's photon) will undergo a state reduction into $|R\rangle$ or $|L\rangle$ state before being sent to the amplifier at whose output one has $G$ photons in the same state as the input. Therefore, the total number of photons entering into Charlie's detector during a time of $t$ second is $GtN_\gamma$, with roughly $GtN_\gamma/2$ photons in each CP state.

Although the angular momentum change (2$\hbar$) of each plate in Charlie's detector caused by an incident $|L\rangle$ photon cancels out that (-2$\hbar$) by a $|R\rangle$ photon, the number of $|L\rangle$ photons is usually unequal to that of $|R\rangle$ photons thanks to quantum fluctuations. The photon number difference between $|L\rangle$ and $|R\rangle$ states is of the order of $\sqrt{tN_\gamma}$ before power amplification, after which time the number difference becomes $G\sqrt{tN_\gamma}$. Hence, the relative angular momentum change of the wave plates can be calculated as of the order of $4\hbar G\sqrt{tN_\gamma}$. One should note that the above estimations for photon number fluctuations suffice for the purpose of this work, albeit they are not accurate because the photons are not in coherent states.

For a disk-like wave plate, the moment of inertia is
\begin{eqnarray}\label{eq:moi}
I_m&=&(\pi/2)\rho D r^4,
\end{eqnarray}
wherein $\rho$ is the mass density, and $D$ stands for the plate thickness with $r$ being the radius of the disk bottom. Then it follows that the variation of the relative angular velocity of the two half-wave plates in Charlie's detector caused by $GtN_\gamma$ CP photons is
\begin{eqnarray}\label{eq:as}
\omega_p&=&\frac{4\hbar G\sqrt{tN_\gamma}}{I_m}=\frac{8\hbar G}{\pi\rho D r^4}\Delta N_\gamma,
\end{eqnarray}
where $\Delta N_\gamma\equiv \sqrt{tN_\gamma}$. Eq. (\ref{eq:as}) shows a linear relationship between $\omega_p$ and $\Delta N_\gamma$, which means that the output of Charlie's detector is determined by the photon number difference between the two CP states detected by Alice at a space-like distance. Therefore, Eq. (\ref{eq:as}) signifies the ``spooky action at a distance", provided that the measurement speed $\propto t^{-1}$ is high enough to satisfy nonlocality condition \cite{Genovese2005}.

From Eq. (\ref{eq:as}), one may estimate an achievable magnitude of $\omega_p$ with existing technologies. Let use a type-I frequency-nondegenerate OPO as a bright source emitting $N_\gamma = 10^{12}$ photon pairs per second (Fig. \ref{fig:exp1}b). The photon pairs are entangled in polarization but separable in wavelength. One photon-bearing beam from the OPO is fed into an optical amplifier with cascaded stages, each may be cavity-enhanced, providing a total gain of $G=10^6$. If the beam wavelength is about $\lambda=1\ \mu$m, then the incident light into Charlie's detector is of the order of $G N_\gamma E_\gamma = 0.2$ W ($E_\gamma$ is the photon energy), the same optical power level as that of Beth's experiment \cite{Beth1936}.

As for the sizes of the wave plates, the lower bound of their thickness is set by the birefringence ($\Delta n$) of the material. If fabricated with some birefringent material akin to KTP, a zero-order half-wave plate of this kind will have a thickness $D = \lambda/(2\Delta n) \approx 5.5 \ \mu$m and a mass density of $3\times 10^3$ kg/m$^3$. For a designed bottom radius of $r = 50 \ \mu$m, one may obtain from Eq. (\ref{eq:as}) roughly $\omega_p \approx 2.5\times10^{-5}$ radius/s within a period of $t = $ 0.1 ms, during which time a signal at the speed of light travels through a distance of about 30 km in the air.

From the above estimations, one may design an experiment on direct observation of the instantaneous influence between entangled photons as follows (Fig. \ref{fig:exp1}a): Twin beams carrying entangled photons are produced by a type-I OPO. One beam is sent to Alice's CP analyzer \cite{Pan2000} with its twin sent into Charlie's detector after being amplified. Alice's analyzer and Charlie's detector are separated by more than 30 km and, if the experiment is ground based, the twin beams may transport via low-noise fibers to arrive their destinations.

During the experiment, the OPO is turned on for a period of $t = $ 0.1 ms to generate entangled photon pairs. Then, the photons received by Alice are projected into $|L\rangle$ or $|R\rangle$ states and, immediately, the photons flying towards Charlie collapse into corresponding CP states according to Eq. (\ref{eq:angular}). The power-amplified beam carrying CP photons enters into Charlie's detector and transmits through the wave plates, whose relative angular speed then changes by a magnitude of $\omega_p$ proportional to $\Delta N_\gamma$ which is recorded by Alice. The photon flux may exceed the detection ability of the photon counters in use, and one may instead utilize quantum-noise-limited heterodyne detectors \cite{xie2018} to measure the light-field amplitudes to obtain $\Delta N_\gamma$. Finally, the OPO is turned off so that the wave plates may freely rotate for a period of say $\tau =$ 300 s during which time the relative rotation angle of the two plates will change by of the order of $\theta=\omega_p \tau\approx 0.4^\circ$, which may be monitored using a polarimeter at 2 $\mu$m where the half-wave plates at 1 $\mu$m become quarter-wave plates.

The above experimental procedure can be repeated as long as allowed by experimental conditions and one may extract the temporal correlation between $\omega_p$ and $\Delta N_\gamma$ from the data,
\begin{eqnarray}\label{eq:corram}
C_p&=&\frac{\sum_i \omega_p(i) \cdot \Delta N_\gamma(i)}{\sqrt{\sum_i |\omega_p(i)|^2} \cdot \sqrt{\sum_i |\Delta N_\gamma(i)|^2}},
\end{eqnarray}
wherein $i$ represents the sequential number of repeated data acquisition in the experiment. Ideally, $\omega_p$ and $\Delta N_\gamma$ should keep their relationship of Eq. (\ref{eq:as}), plugging which into Eq. (\ref{eq:corram}) leads to $C_p = 1$. Therefore, an experimental value of $C_p$ approaching unity is a signature of the ``spooky action at a distance". Otherwise, one should have $C_p = 0$.

Next we turn to VLI observation, for which the observer must be in a inertial frame moving at a velocity $v$ relative to the lab frame with $v$ satisfying the inequality (\ref{eq:VLIcon}). The amplitude of $|t_3-t_1|$ is primarily determined by the speed of state collapse in Alice's analyzer. To be conservative, one may assume $|t_3-t_1| \approx 1$ ps and then from Eq. (\ref{eq:VLIcon}) obtain $|v| > 3$  m/s for VLI observation if $|x_3-x_1| \ge $ 30 km. So, if the observer (David) moves at $v > 3$ m/s along the propagation direction of Charlie's photon, from Eq. (\ref{eq:timing}) it follows that $t'_3-t'_1 < $ 0. In other words, for each photon pair, Charlie's photon will arrive at his detector before Alice's photon is received by her analyzer, in which case Charlie's photon carries no angular momentum when entering into his detector that outputs $\theta =0$.

In contrast, if the observer (Frank) moves at $v < -3 $  m/s, he will observe according to Eq. (\ref{eq:timing}) that $t'_3-t'_1 >$ 0, i.e., Charlie's detector receives a photon later than Alice's analyzer and, consequently, the detection of Alice's photon in a CP state will instantaneously reduce Charlie's photon into another CP state before it enters into his detector. This photon carrying nonzero angular momentum will exert a mechanical torque onto each have-wave plate in Charlie's detector that, as a result, outputs $\theta \approx 0.4^\circ$. The discrepancy between the observations of David and Frank moving in opposite directions signifies VLI.

To conclude, we have studied direct observation of the instantaneous influence between polarization entangled photons. The key idea is the mechanical detection of the CP state of a photon that carries angular momentum and exerts a torque on a half-wave plate whose mechanical motion state varies depending on the specific CP state of the incident photon. Because no local realist model is assumed in the investigation, if implemented, the stated observation should be a model independent evidence for quantum nonlocality. Moreover, we have firmly established the intrinsic connection between quantum nonlocality and Lorentz invariance violation, which indicates fundamental physics beyond the Standard Model.

\begin{acknowledgments}
B.X. and P.Y. made equal contributions to this work, of which S.F. conceived the core ideas and which was partially supported by the National Natural Science Foundation of China (grant No. 51575164).
\end{acknowledgments}


\providecommand{\noopsort}[1]{}\providecommand{\singleletter}[1]{#1}%
%

\end{document}